\documentclass[journal]{IEEEtran}
\usepackage{graphicx,amssymb,mathrsfs,amsmath,array,color}
\usepackage{subfigure}
\usepackage{multirow}
\usepackage{booktabs}
\usepackage{cite}
\usepackage[colorlinks,urlcolor=blue,driverfallback=dvipdfm]{hyperref}
\usepackage{bm} 
\usepackage{makecell} 
\usepackage{pifont}
\newcommand{\cmark}{\ding{51}}
\newcommand{\xmark}{\ding{55}}

\usepackage{CJKutf8}

\begin{document}
\title{Dual-Branch Subpixel-Guided Network for Hyperspectral Image Classification}
\begin{CJK}{UTF8}{gbsn}

\author{Zhu Han,~\IEEEmembership{Student Member,~IEEE,}
        Jin Yang,
        Lianru Gao,~\IEEEmembership{Senior Member,~IEEE,}
        Zhiqiang Zeng,~\IEEEmembership{Student Member,~IEEE,}
        Bing Zhang,~\IEEEmembership{Fellow,~IEEE,}
        and Jocelyn Chanussot,~\IEEEmembership{Fellow,~IEEE}

\thanks{This work was supported by the National Natural Science Foundation of China under Grant 42325104 and Grant 62161160336. (\emph{Corresponding author: Lianru Gao.})}
\thanks{Z. Han is with the Key Laboratory of Digital Earth Science, Aerospace Information Research Institute, Chinese Academy of Sciences, Beijing 100094, China, and with the International Research Center of Big Data for Sustainable Development Goals, Beijing 100094, China, and also with the College of Resources and Environment, University of Chinese Academy of Sciences, Beijing 100049, China (e-mail: hanzhu19@mails.ucas.ac.cn).}
\thanks{J. Yang and L. Gao is with the Key Laboratory of Computational Optical Imaging Technology, Aerospace Information Research Institute, Chinese Academy of Sciences, Beijing 100094, China (e-mail: raemyj@163.com; gaolr@aircas.ac.cn).}
\thanks{Z. Zeng is with the Beijing Institute of Remote Sensing Equipment, Beijing 100854, China (e-mail: zengzq@buaa.edu.cn).}
\thanks{B. Zhang is with the Aerospace Information Research Institute, Chinese Academy of Sciences, Beijing 100094, China, and also with the College of Resources and Environment, University of Chinese Academy of Sciences, Beijing 100049, China (e-mail: zhangbing@aircas.ac.cn).}
\thanks{J. Chanussot is with the Univ. Grenoble Alpes, CNRS, Grenoble INP, IJK, Grenoble 38000, France, and also with the Aerospace Information Research Institute, Chinese Academy of Sciences, 100094 Beijing, China (e-mail: jocelyn.chanussot@inria.fr).}
}

\markboth{Accepted by IEEE Transactions on Geoscience and Remote Sensing,~Vol.~XX, No.~XX, ~XXXX,~2024}
{Han \MakeLowercase{\textit{et al.}}: }

\maketitle
\begin{abstract}
Deep learning (DL) has been widely applied into hyperspectral image (HSI) classification owing to its promising feature learning and representation capabilities. However, limited by the spatial resolution of sensors, existing DL-based classification approaches mainly focus on pixel-level spectral and spatial information extraction through complex network architecture design, while ignoring the existence of mixed pixels in actual scenarios. To tackle this difficulty, we propose a novel dual-branch subpixel-guided network for HSI classification, called DSNet, which automatically integrates subpixel information and convolutional class features by introducing a deep autoencoder unmixing architecture to enhance classification performance. DSNet is capable of fully considering physically nonlinear properties within subpixels and adaptively generating diagnostic abundances in an unsupervised manner to achieve more reliable decision boundaries for class label distributions. The subpixel fusion module is designed to ensure high-quality information fusion across pixel and subpixel features, further promoting stable joint classification. Experimental results on three benchmark datasets demonstrate the effectiveness and superiority of DSNet compared with state-of-the-art DL-based HSI classification approaches. The codes will be available at \url{https://github.com/hanzhu97702/DSNet}, contributing to the remote sensing community.

\end{abstract}
\graphicspath{{figures/}}


\section{Introduction}
\IEEEPARstart{H}{yperspectral} imaging is a technique for exploring the spectral properties of ground targets with the fine resolution of scene radiance \cite{bioucas2013challenge}. Owing to its rich spectral characteristics, each pixel in hyperspectral images (HSIs) can be regarded as an approximately continuous spectral curve, enabling various materials to be effectively identified and discriminated \cite{8101519, 10475370}. Benefited from inherent values within the cubic data architecture, HSI has been widely utilized in many fields, including precision agriculture\cite{SINGH2020121, HABOUDANE2004337}, urban planing\cite{srivastava2019understanding, qiu2019local}, target detection \cite{gao2023bs, 10379605} and mineral monitoring\cite{1220247, peyghambari2021hyperspectral}. As one of the primary HSI processing technologies, HSI classification is an essential foundation and aims at generating high-precision classification maps that reflect the ground distribution information.

In the past few decades, numerous traditional machine learning methods have be proposed for HSI classification, such as the support vector machine (SVM)\cite{9206124}, random forest (RF)\cite{9781430}, and the Bayesian model\cite{8978952}. Nevertheless, these traditional methods cannot consider sufficient spectral-spatial features to establish the relationship between pixels in the spatial dimension. A large number of advanced dimensionality reduction and feature extraction methods have been further introduced to improve the classification accuracy, including subspace learning\cite{gao2014subspace, fu2020learning}, principal component analysis (PCA) \cite{kang2017pca, uddin2021pca}, morphological profiles\cite{fauvel2008spectral, dalla2010classification, 10521690}, Gabor filters \cite{li2014gabor, he2016discriminative}, ensemble learning\cite{8948304} and superpixel-based analysis\cite{fang2015spectral, sun2021spassa}. However, the extracted features by these above-mentioned approaches rely on manual design by experts and tend to be shallow, making the recognition and robustness in complex scenarios unsatisfactory.

With the explosive growth of computer vision and artificial intelligence theory, deep learning (DL) solves this bottleneck and can automatically extract robust and high-level representations from a data-driven perspective\cite{li2019deep, 9780199, ZENG2023242, han2023spatio, 10504844}. The most prominent DL-based HSI classification methods are convolutional neural networks (CNNs)\cite{lee2017going, zhang2018diverse, yu2021feedback}, graph convolutional networks (GCNs)\cite{wan2019multiscale, hong2020graph, mou2020nonlocal}, recurrent neural networks (RNNs)\cite{mou2017deep, zhang2018spatial, pan2020spectral} and Transformers\cite{hong2021spectralformer, he2021spatial}, which explores the spectral and spatial feature representations from different views. Specifically, CNNs aim at extracting spatial contextual representations by employing two-dimensional (2-D) or three-dimensional (3-D) convolution kernels, whereas the fixed receptive field limits the model learning ability to retrieve large-scale contextual information. GCNs and RNNs can capture topological associations among samples and model spectral sequences in the HSI, respectively. Following the adoption of the self-attention mechanism, Transformer-based methods can fully learn global spectral-spatial structure information, but the quadratic complexity limits its computational efficiency. Furthermore, recent studies have demonstrated that the hybrid architecture of these DL models can solve the imperfection of feature exploration and greatly improve the performance of HSI classification in complex scenes \cite{roy2021morphological, sun2022spectral, dong2022weighted}. Although existing DL-based approaches have achieved great success in the field of HSI classification, these methods usually treat each pixel as a pure spectrum for classifier training and ignore the existence of mixed pixels owing to the impact of low spatial resolution in actual scenarios. In general, mixed pixels reflect the spectral mixing characteristics of different ground objects in the scene, but existing studies fail to consider the subpixel information that characterizes the spatial distribution of ground objects within a single pixel, which inevitably leads to misclassification by the classifier. In addition, subpixel in the HSI usually has certain physical mixing constraints\cite{1000320, keshava2002spectral}, thus how to naturally integrate information provided by discriminative classifier and subpixel information to improve classification performance still faces great challenges.

\begin{figure}[!t]
	  \centering
		\subfigure[]{
			\includegraphics[width=0.5\textwidth]{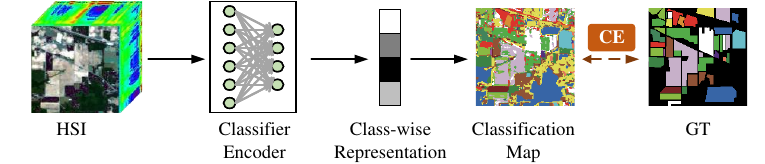}
            \label{fig:DL classification}
		}\qquad
		\subfigure[]{
			\includegraphics[width=0.5\textwidth]{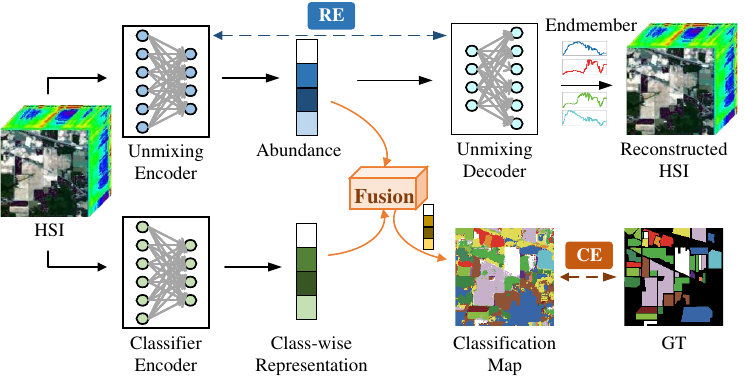}
            \label{fig:DL class_unmix}
		}\qquad
         \caption{Illustration to clarify the similarities and differences between the existing HSI classification method and the subpixel-guided HSI classification method using DL. (a) Workflow for the existing DL-based HSI classification method. (b) Workflow for the subpixel-guided HSI classification method.}
\label{fig:modal difference}
\end{figure}

The rise of hyperspectral unmixing (HU) technology supports the realization of intelligent subpixel interpretation. HU deals with this challenging issue by separating the mixed pixel into a set of endmember signatures and their corresponding fractional abundances\cite{ren2022hyperspectral,9865216}. Depending on geometry, spectral reflection and refraction characteristics, HU approaches mainly rely on two mixing assumptions to describe photo interactions underlying the observations: linear mixing model (LMM) and nonlinear mixing model (NLMM) \cite{fan2009comparative}. LMM considers the macroscopic mixing scale and assumes the linear combination of different endmember spectra in the scene \cite{dobigeon2008semi}. NLMM introduces the multiple light scatterings in the microscopic scale and achieves a suitable approximation to the actual mixing \cite{dobigeon2013nonlinear,heylen2014review}. Nevertheless, NLMM usually needs a large number of reliable metrics and prior knowledge, which is difficult to be calculated in real scenarios \cite{altmann2012supervised, heylen2015multilinear, yang2018band}. In recent decades, DL-based unmixing methods have gradually become the mainstream for handling mixed pixels in the HSI. As an important representative of DL, autoencoder (AE) can achieve blind HU by imposing certain abundance and endmember constraints with an encoder-decoder architecture, which has been proven to be effective in the field of HU\cite{su2019daen, wang2019nonlinear, han2020deep, palsson2020convolutional,zhao2021plug, gao2021cycu, han2022autonas, han2022multimodal}. Nevertheless, it should be noted that existing HSI classification and unmixing methods based on DL are conventionally performed independently, lacking a robust and effective connection. For some hyperspectral satellite data with low spatial resolution, such as Airborne Visible/Infrared Imaging Spectrometer (AVIRIS), Environmental Mapping and Analysis Program (EnMAP) or Gaofen-5 (GF-5) satellite, there are a large number of mixed pixels in the scene, which inevitably degrades the subsequent per-pixel fine classification processing\cite{ye2020application, zhong2021advances, lekka2024appraisal}. Only relying on increasing the complexity of DL-based HSI classification models to enhance the modeling ability cannot be applied in actual emergency scenarios due to high training time and model parameters. Although some fusion-based methods have been proposed as a prepossessing technique to promote classification performance, the prior knowledge of other remote sensing data sources is required to assist model learning and the adopted deterministic degradation modeling cannot handle realistically-blurred HSIs\cite{he2021cnn, 10197521, 10233913}, which further affects the application of subsequent downstream tasks. Potential subpixel information is often ignored and not fully mined into existing DL-based classification approaches. Therefore, it is necessary to explore an united and intelligent data-driven framework for both HSI classification and unmixing tasks rather than being implemented step-wise. Furthermore, introducing additional subpixel information into DL models usually leads to information redundancy and underutilization, so exploring the efficient fusion strategy during the joint optimization procedure can facilitate better interpretation of class label distribution.

Motivated by the above concerns, we propose a dual-branch subpixel-guided network for HSI classification, called DSNet, in which a deep AE unmixing architecture with physically nonlinear properties is incorporated into the CNN-based classifier network, to effectively fuse subpixel and class-wise representations of HSI to enhance classification performance. Fig. \ref{fig:modal difference} briefly illustrates the similarities and differences between the existing HSI classification method and the proposed subpixel-guided HSI classification method using DL. Different from previous DL-based HSI classification studies that generally focus on pixel-level spectral and spatial information extraction through complex network architecture design, this is the first attempt to adopt the reconstruction (RE) loss and cross-entropy (CE) loss for DSNet training, to collaboratively explore latent correlations from subpixel and pixel information in the HSI. The proposed subpixel-guided HSI classification method can mine inherent spatial distribution of different materials from their corresponding spectral attributes to improve the discrimination ability of the classifier. For this purpose, the subpixel fusion module is designed to aggregate diagnostic abundances and class-wise representations to further guide the classifier network toward a more accurate classification direction. In brief, the major contributions of this paper can be summarized as follows.




\begin{itemize}
    \item We propose a subpixel-guided deep network by introducing a deep AE unmixing architecture for HSI classification tasks, called DSNet. DSNet is capable of estimating subpixel-level abundances and generating discriminative class-wise representations more automatically and efficiently, thereby yielding a significant classification performance improvement.
    \item The deep AE unmixing architecture considers a general unmixing modeling consisting of a linear mixture component and a physically nonlinear mixture component, which provides a complete and physically meaningful subpixel prior information for the CNN-based classifier network.
    \item The subpixel fusion module is developed to ensure high-quality information fusion across pixel and subpixel features, which further achieves stable joint classification and facilitates a better separation of different classes.
\end{itemize}

\begin{figure*}[!t]
	  \centering
		\subfigure{
			\includegraphics[width=1\textwidth]{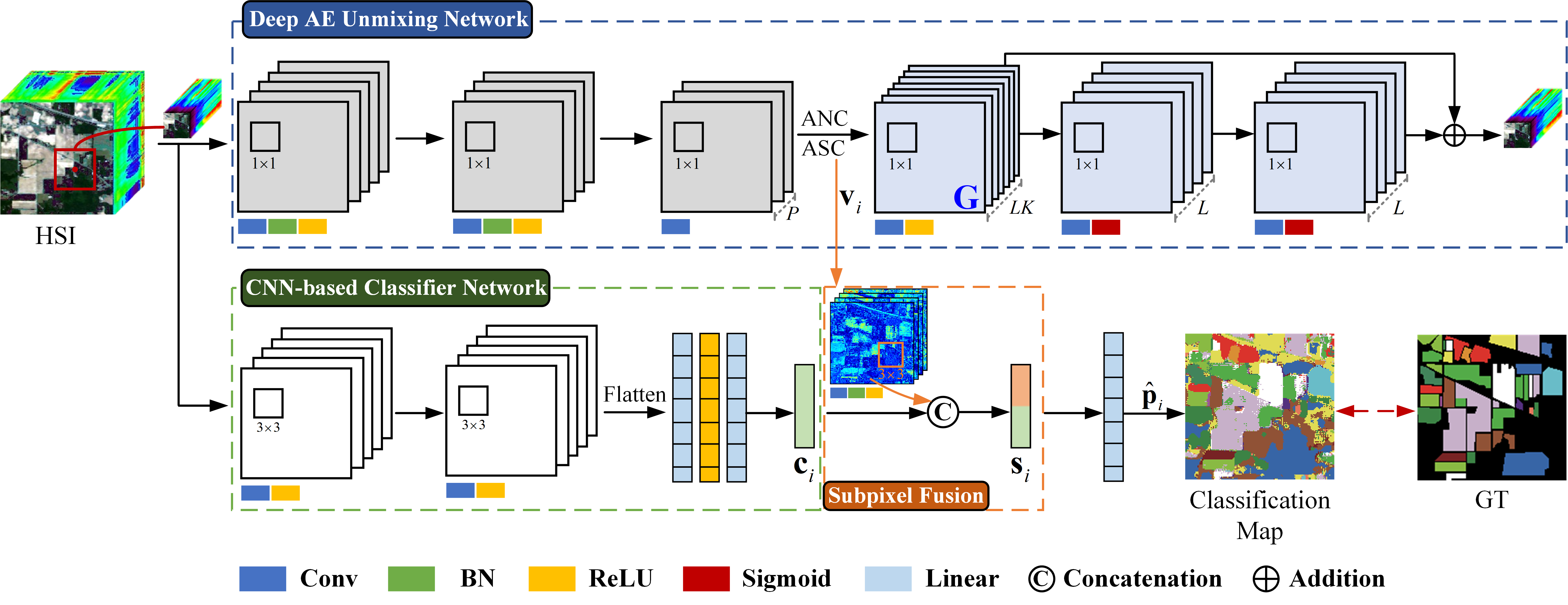}
		}
        \caption{The framework of the proposed DSNet, including deep AE unmixing network, CNN-based classifier network and subpixel fusion module. The deep AE unmixing network is designed by considering a general mixing decoder with physically nonlinear constraints, and further extract useful subpixel-level abundance information from the HSI in an unsupervised manner. The CNN-based classifier network extracts the spectral-spatial information within the HSI to obtain pixel-level class features. The subpixel fusion module aims at integrating the abundance information and class features to ensure high-quality information fusion and enhance model representation capability.}
\label{fig:workflow}
\end{figure*}

The remaining part of this paper is organized as follows. Section II elaborates the implementation details of our DSNet. Section III presents the extensive experiments and analyses on three HSI benchmark datasets. Finally, conclusions are drawn and summarized in Section IV.

\section{Problem Formulation and Method}
In this section, we start with a review of the existing AE-based unmixing approaches, and then provide a detailed description of the proposed DSNet method.
\subsection{Brief Review of AE-based Unmixing Approaches}
As a simple and commonly used spectral mixing model, the LMM assumes that a given spectral vector of $i$-th pixel in the HSI is generated by the linear combination of different endmember spectra and their corresponding abundances, which can be formulated as follows:
\begin{equation}
\label{eq1}
    \mathbf{y}_{i} =\mathbf{M}\mathbf{a}_{i}+\mathbf{n}_{i}
\end{equation}
where $\mathbf{y}_{i}\in \mathbb{R}^{L}$ represents the input spectral vector with $L$ spectral bands. $\mathbf{M}=[\mathbf{m}_{1},\mathbf{m}_{2},\cdots,\mathbf{m}_{P}]\in \mathbb{R}^{L\times P}$ is the endmember matrix with $P$ endmember categories and $\mathbf{m}_{i}$ denotes the $i$-th endmember vector. $\mathbf{a}_{i}\in \mathbb{R}^{P}$ is the fractional abundance vector for different endmembers in the $i$-th observed pixel. $\mathbf{n}_{i}\in \mathbb{R}^{L}$ denotes the noise vector in the HSI.

Under the matrix notation, we can rewrite (\ref{eq1}) in a compact matrix form as follows:
\begin{equation}
\label{eq2}
    \mathbf{Y}=\mathbf{M}\mathbf{A}+\mathbf{N}
\end{equation}
where $\mathbf{Y}\in \mathbb{R}^{L\times N}$ and $\mathbf{N}\in \mathbb{R}^{L\times N}$ denote the input HSI with $N$ pixels and the noise matrix, respectively. $\mathbf{A}=[\mathbf{a}_{1},\mathbf{a}_{2},\cdots,\mathbf{a}_{N}] \in \mathbb{R}^{P\times N}$ is the abundance matrix, where each abundance vector $\mathbf{a}_{i}$ should satisfy the abundance sum-to-one constraint (ASC) and the abundance non-negativity constraint (ANC), that is:
\begin{equation}
\label{eq3}
    \mathbf{1}_{P}^{T}\mathbf{A}=\mathbf{1}_{N}^{T}
\end{equation}
\begin{equation}
\label{eq4}
    \mathbf{A}\ge\mathbf{0}
\end{equation}

Furthermore, due to the existence of multiple scattering interactions between different materials, NLMM is proposed to delineate this complex mixture mechanism. The general form of NLMM is regarded as the sum of a linear mixture and a nonlinear fluctuation that can be parameterized by fractional abundances and endmembers\cite{chen2012nonlinear, wang2019nonlinear}. Its mathematical formulation can be written as
\begin{equation}
\label{eq5}
    \mathbf{Y}=\mathbf{M}\mathbf{A}+\mathbf{\Phi}(\mathbf{M}\mathbf{A})+\mathbf{N}
\end{equation}
where $\mathbf{\Phi}$ is the nonlinear mapping function applied to the linear transform $\mathbf{MA}$. To achieve an adaptive spectral mixing mechanism from a data-driven perspective, DL-based methods are widely applied in the field of HU. As a typical unsupervised method of DL, AE has been applied to model the spectral mixture process owing to its powerful learning and reconstruction capability. In general, the AE consists of two parts, namely, an encoder and a decoder.
\subsubsection{Encoder}
The encoder part is designed to transform the input pixel $\mathbf{y}_{i}$ into a hidden representation $\mathbf{v}_{i}$ by utilizing some trainable network parameters, which can be written as
\begin{equation}
\label{eq6}
    \mathbf{v}_{i}=f_{E}(\mathbf{y}_{i})=f(\mathbf{W}^{(e)}\mathbf{y}_{i}+\mathbf{b}^{(e)}),
\end{equation}
where $f(\cdot)$ is the nonlinear activation function, such as the rectified linear unit (ReLU), the leaky ReLU (LReLU) and the sigmoid function. $\mathbf{W}^{(e)}$ and $\mathbf{b}^{(e)}$ denote the weight and bias in the $e$-th encoder part.

\subsubsection{Decoder}
Based on different mixing assumptions, the decoder part employs one or more hidden layers to reconstruct the input  pixel from the hidden representation $\mathbf{v}_{i}$ and the reconstruction process is expressed as
\begin{equation}
\label{eq7}
    \mathbf{\hat{y}}_{i}=f_{D}(\mathbf{v}_{i})=\mathbf{W}^{(d_{1})}\mathbf{v}_{i}+\mathbf{\Phi}(\mathbf{W}^{(d)}\mathbf{v}_{i})
\end{equation}
where $\mathbf{\hat{y}}_{i}\in \mathbb{R}^{L}$ is the reconstructed spectral vector. $\mathbf{W}^{(d_{1})}$ and $\mathbf{W}^{(d)}$ represent the weight matrix in the first 
 and other $d$-th decoder part. $\mathbf{\Phi}(\cdot)$ is the nonlinear interaction among the endmembers. When the observed scene is based on LMM, the decoder part has one hidden layer and $f_{D}(\mathbf{v}_{i})=\mathbf{W}^{(d_{1})}\mathbf{v}_{i}$. Since the structure of the decoder satisfies the spectral mixing process, the results of the extracted abundance vector $\mathbf{\hat{a}}_{i}$ and endmember matrix $\mathbf{\hat{M}}$ in the AE can be regarded as $\mathbf{v}_{i}$ and $\mathbf{W}^{(d_{1})}$, respectively.

\subsection{Dual-Branch Subpixel-Guided Network}
To effectively aggregate diagnostic subpixel information for HSI classification tasks, the proposed DSNet method is used to introduce a deep AE unmixing architecture with physically nonlinear properties to fully explore inherent spectral and spatial correlations in the HSI. As illustrated in Fig. \ref{fig:workflow}, our DSNet is composed of a deep AE unmixing network and CNN-based classifier network, in which the subpixel fusion module is designed to ensure efficient utilization of different levels of discriminative information. The overall network configuration in the proposed DSNet is shown in Table \ref{tab:network architecture}.

\begin{table}[!t]
\centering
\caption{Network configuration of the proposed DSNet.}
\resizebox{0.5\textwidth}{!}{
    \begin{tabular}{c|c|cc|cc}
    \toprule[1.5pt]
    \multirow{2}{*}{Architecture} & \multirow{2}{*}{Pathway} & \multicolumn{2}{c|}{Layer composition} & \multicolumn{2}{c}{Unit} \\
    \cline{3-6} & & unmixing & classifier & unmixing & classifier \\
    \hline
    \multirow{9}{*}{Dual-Branch Encoder} & \multirow{3}{*}{Block 1} & Conv 1$\times$1 & Conv 3$\times$3 & \multirow{3}{*}{$L$ / 2} & \multirow{3}{*}{64}\\
    ~ & ~ & BN & ReLU & ~ & ~\\
    ~ & ~ & ReLU & ~ & ~ & ~\\
    \cline{2-6}
    ~ & \multirow{3}{*}{Block 2} & Conv 1$\times$1 & Conv 3$\times$3 & \multirow{3}{*}{$L$ / 4} & \multirow{3}{*}{100}\\
    ~ & ~ & BN & ReLU & ~ & ~\\
    ~ & ~ & ReLU & Flatten & ~ & ~\\
    \cline{2-6}
    ~ & \multirow{3}{*}{Block 3} & Conv 1$\times$1 & Linear & \multirow{2}{*}{$P$} & \multirow{2}{*}{100}\\
    ~ & ~ & ANC & ReLU & ~ & ~\\
    ~ & ~ & ASC & Linear & ~ & $P$\\
    \cline{2-6}
    \hline
    \multirow{4}{*}{Decoder $\&$ Classifier Output} & \multirow{2}{*}{Block 4} & Conv 1$\times$1 & Subpixel Fusion &  \multirow{2}{*}{$LK$} & \multirow{2}{*}{$S$}\\
    ~ & ~ & ReLU & ~ & ~ & ~\\
    \cline{2-6}
    ~ & \multirow{4}{*}{Block 5} & Conv 1$\times$1 & Linear & \multirow{2}{*}{$L$} & \multirow{2}{*}{$P$}\\
    ~ & ~ & Sigmoid & ~ & ~\\
    ~ & ~ & Conv 1$\times$1 & ~ & \multirow{2}{*}{$L$}\\
    ~ & ~ & Sigmoid & ~ & ~\\
    \bottomrule[1.5pt]
    \end{tabular}
}
\label{tab:network architecture}
\end{table}
 
\subsubsection{Deep AE Unmixing Network}
The structure of the deep AE unmixing network aims at extracting subpixel-level abundance information by building a general mixing decoder with physically nonlinear constraints during the reconstruction process. As illustrated in Table \ref{tab:network architecture}, block 1-3 represent the encoder part in the deep AE unmixing architecture, and block 4-5 are the general mixing decoder part. In essence, deep AE unmixing network can realize blind unmixing process from HSIs and extract useful abundance information as internal subpixel knowledge for the classifier. Given the input patch $\left\{\mathbf{x}_{i}\right\}_{i=1}^{N_{S}}\in \mathbb{R}^{L\times H\times H}$ with the $H\times H$ spatial size and $N_{S}$ training samples, the encoder part of AE is served as a feature extractor to map the input into high-dimensional abundance representations by the following transformation:
\begin{equation}
\label{eq8}
\begin{aligned}
    \mathbf{h}_{i}^{(e)} = \left\{
    \begin{array}{ll}
    f(BN_{\gamma,\beta}(\mathbf{W}_{h}^{(e)}\mathbf{x}_{i}+\mathbf{b}_{h}^{(e)})), & e = 1 \\
    f(BN_{\gamma,\beta}(\mathbf{W}_{h}^{(e)}\mathbf{h}_{i}^{(e-1)}+\mathbf{b}_{h}^{(e)})), & e = 2 \\
    \mathbf{W}_{h}^{(e)}\mathbf{h}_{i}^{(e-1)}+\mathbf{b}_{h}^{(e)},  & e = 3
    \end{array}
    \right.
\end{aligned}
\end{equation}
where $\mathbf{h}_{i}^{(e)}\in \mathbb{R}^{P\times H\times H}$ denotes the encoded hierarchical representation of HSI data in the $e$-th encoder layer. $f(\cdot)$ is the ReLU activation function. $BN_{\gamma,\beta}(\mathbf{x}_{i}) = \gamma\mathbf{\hat{x}}_{i} + \beta$ represents the batch normalization (BN) layer to speed up the parameter learning and avoid the problem of vanishing gradients in the training phase. $\left\{\mathbf{W}_{h}^{(e)},\mathbf{b}_{h}^{(e)}\right\}$ is the set of the corresponding weight and bias matrix in the 1 $\times$ 1 convolution operation. It is emphasized that 1 $\times$ 1 convolution is equivalent to the fully connected structure in the traditional AE network, which can further ensure the integrity of the spatial structure for the extracted abundance maps and fully explore the spectral relationship between different ground objects.

To guarantee ANC and ASC constraints, the absolute value rectification and summed normalization are adopted to the encoded hierarchical representation $\mathbf{h}_{i}^{(3)}$, which can be written as
\begin{equation}
\label{eq9}
    \mathbf{v}_{i}=\frac{|\mathbf{h}_{i}^{(3)}|}{\sum_{i=1}^{P}|\mathbf{h}_{i}^{(3)}|}
\end{equation}
where $\mathbf{v}_{i}\in \mathbb{R}^{P\times H\times H}$ represents the extracted abundance results from the encoder part.

Unlike previous works that set the fixed number of decoder layers for unmixing, the proposed general mixing decoder is designed to reconstruct the input with different network architectures and takes into account the number of decoder layers $K$ on unmixing performance. Based on (\ref{eq5}) and (\ref{eq7}), a novel weight matrix $\mathbf{G}\in \mathbb{R}^{LK\times P}$ is proposed to unify and simplify $\mathbf{W}^{(d_{1})}$ and $\mathbf{W}^{(d)}$, so that the unmixing performance is insensitive to the number of decoder layers while satisfying physically nonlinear properties, denoted as follows:
\begin{equation}
\label{eq10}
\mathbf{W}^{(d_{1})}=\sum_{k=1}^{K}\mathbf{G}_{k}
\end{equation}
\begin{equation}
\label{eq11}
\mathbf{W}^{(d)}=\mathcal{T}_{LK\rightarrow L}(\mathbf{G})
\end{equation}
where $\mathbf{G}_{k}\in \mathbb{R}^{L\times P}$ is the each layer element in $\mathbf{G}$. $\mathbf{W}^{(d_{1})}$ can be regarded as the joint interaction of each layer in the decoder. $\mathcal{T}_{LK\rightarrow L}$ is the nonlinear mapping function that transforms features from the $LK$ dimension space to the $L$ dimension space. Then, the reconstruction process of the general mixing decoder can be rewritten as
\begin{equation}
\label{eq12}
\mathbf{\hat{x}}_{i}=f_{D}(\mathbf{v}_{i})=\sum_{k=1}^{K}\mathbf{G}_{k}\mathbf{v}_{i}+\mathcal{T}_{LK\rightarrow L}(\mathbf{G}\mathbf{v}_{i})
\end{equation}
where $\mathbf{\hat{x}}_{i}\in \mathbb{R}^{L\times H\times H}$ is the reconstructed patch obtained from the deep AE unmixing network. By effectively designing a single weight matrix $\mathbf{G}$, the proposed general mixing decoder can guarantee lower computational complexity and achieve better physical interpretability in practical scenarios. More precisely, block 4 corresponds to the front part in (\ref{eq12}) and describes the LMM mechanism while maintaining output non-negativity with the help of the ReLU activation function. Block 5 represents nonlinear interactions between different materials by applying a combination of 1 $\times$ 1 convolution operator and sigmoid activation function with the bias setting.

\subsubsection{CNN-based Classifier Network}
The spectral-spatial information within the HSI is considered and extracted by the CNN-based classifier network to obtain pixel-level class features $\mathbf{c}_{i}\in \mathbb{R}^{P}$. In this paper, we only adopt a simple 2-D convolution architecture to extract the spectral-spatial information of HSIs, which can better reflect the advantage of introducing subpixel information. As illustrated in Table \ref{tab:network architecture}, two 3 $\times$ 3 convolution layers are initially deployed to extract feature maps with rich spatial information, followed by the ReLU activation function. Then, the extracted features are flattened and fed into two linear layers to output the pixel-level class feature $\mathbf{c}_{i}$. $\mathbf{c}_{i}$ can provide the discriminative class probabilities for DSNet to aid subsequent training of the subpixel fusion module. This process can be expressed as
\begin{equation}
\label{eq13}
\mathbf{c}_{i}=f_{CNN}(\mathbf{x}_{i})
\end{equation}
where $f_{CNN}(\cdot)$ denotes the transformation process in the CNN-based classifier network.

\subsubsection{Subpixel Fusion Module}
To achieve efficient combination of subpixel and pixel information, the subpixel fusion module is proposed to generate discriminative class-wise representations for the supervised training of DSNet. Given the input abundance patch $\mathbf{v}_{i}$ and class feature $\mathbf{c}_{i}$, due to the difference of the spatial dimension, the front part of subpixel fusion module is designed to maintain the consistency of spatial size by converting the abundance patch into one-dimension space through one 3 $\times$ 3 convolution layer with a stride of 2. Then, the concatenation operation is adopted to fuse the transformed abundance and class feature, so that the integrity of abundance information is preserved as much as possible in the training process. The formula of the subpixel fusion module can be summarized as
\begin{equation}
\label{eq14}
\mathbf{s}_{i}=\lbrack{\rm Flatten}(f(BN_{\gamma,\beta}(\mathbf{W}_{sub}\mathbf{v}_{i}+\mathbf{b}_{sub}))), \mathbf{c}_{i}\rbrack
\end{equation}
where $\mathbf{s}_{i}\in \mathbb{R}^{S}$ is the fused joint representation derived from the subpixel fusion module. $[\cdot]$ and $\rm Flatten(\cdot)$ stand for concatenation and flatten operation, respectively. $\mathbf{W}_{sub}$ and $\mathbf{b}_{sub}$ denote the weight and bias matrix for spatial dimension reduction in the 3 $\times$ 3 convolution layer.

Finally, the class prediction is output by a linear layer that transforms the fused joint representation into the dimension of the class, and the process can be formulated as
\begin{equation}
\label{eq15}
\mathbf{\hat{p}}_{i} = \mathbf{W}_{out}\mathbf{s}_{i}+\mathbf{b}_{out}
\end{equation}
where $\mathbf{\hat{p}}_{i}\in \mathbb{R}^{P}$ denotes the output class prediction by DSNet. $\mathbf{W}_{out}$ and $\mathbf{b}_{out}$ represent the weight and bias matrix for transformation in the linear layer of Block 5.

\subsection{Objective Function}
As stated before, the objective function of the proposed DSNet is realized by two phases. One phase is to train the deep AE unmixing network by minimizing the RE loss based on spectral angle distance (SAD), given by
\begin{equation}
\label{eq16}
    L_{\text{RE}}=\frac{1}{N_{S}}\sum_{i=1}^{N_{S}}{\rm arccos}\left(\frac{\mathbf{\hat{x}}_{i}^{T}\mathbf{x}_{i}}{\left\|\mathbf{\hat{x}}_{i}\right\|_{2}\left\|\mathbf{x}_{i}\right\|_{2}}\right).
\end{equation}
where $N_{S}$ represents the number of training samples. $\mathbf{x}_{i}$ and $\mathbf{\hat{x}}_{i}$ denote the $i$-th patch in the input HSI and the reconstructed HSI, respectively.

The second phase aims at training the CNN-based classifier network by adopting the CE loss between the fused class-wise representation and true label in the ground truth (GT). The standard CE loss is calculated as 
\begin{equation}
\label{eq17}
    L_{\text{CE}}=-\frac{1}{N_{S}}\sum_{i=1}^{N_{S}}\mathbf{p}_{i}\log(\mathbf{\hat{p}}_{i})
\end{equation}
where $\mathbf{p}_{i}$ and $\mathbf{\hat{p}}_{i}$ are the one-hot encoding of the true label and the corresponding class prediction of DSNet.

The overall loss of DSNet can be formulated as
\begin{equation}
\label{eq18}
    L= \lambda L_{\text{RE}} + (1-\lambda) L_{\text{CE}}
\end{equation}
where $\lambda$ is a hyperparameter to balance different objective functions, and the range of $\lambda$ is [0, 1]. 

\begin{table*}[!t]
\centering
\caption{Class-Specific and Overall Classification Accuracy (\%) of Different Methods on the Indian Pines Dataset}
\resizebox{1\textwidth}{!}{
\begin{tabular}{c||c|c|c|c|c|c|c|c|c|c}
\hline \hline
\multirow{2}{*}{Class Name} & \multirow{2}{*}{\makecell{Training\\Samples}} & \multirow{2}{*}{\makecell{Test\\Samples}} & \multicolumn{4}{c|}{Classic Backbone Networks} & \multicolumn{4}{c}{Hybrid Backbone Networks}\\
\cline{4-11}
~ & ~ & ~ & 2-D CNN\cite{paoletti2019deep}  & 3-D CNN\cite{chen2016deep}  & GRU\cite{pan2020spectral}  & ViT\cite{vaswani2017attention}  & MorphConv\cite{roy2021morphological}  & SSFTT\cite{sun2022spectral}  & WFCG\cite{dong2022weighted}  & DSNet \\
\hline
Corn Notill & 50 & 1384 & 77.24 & 65.25 & 72.33 & 66.55 & 71.53 & \bf 85.62 & 80.92 & 83.96\\
Corn Mintill & 50 & 784 & 84.57 & 59.57 & 83.16 & 88.52 & 84.57 & 90.43 & 90.69 & \bf 94.90\\
Corn & 50 & 184 & 92.39 & 94.57 & 73.37 & 88.59 & 97.83 & 97.28 & \bf 100.00 & 99.46\\
Grass Pasture & 50 & 447 & 94.63 & 89.71 & 87.02 & 97.09 & 93.74 & 90.60 & 94.41 & \bf 98.21\\
Grass Trees & 50 & 697 & 78.62 & 95.12 & 81.35 & 85.65 & 92.25 & 95.95 & \bf 99.43 & 97.27\\
Hay Windrowed & 50 & 439 & 92.26 & 98.63 & 94.76 & 98.86 & 98.86 & 98.63 & \bf 100.00 & 99.54\\
Soybean Notill & 50 & 918 & 79.63 & 79.74 & 84.86 & 93.03 & 73.31 & 85.51 & 82.79 & \bf 95.10\\
Soybean Mintill & 50 & 2418 & 67.74 & 67.70 & 66.50 & 73.95 & 56.95 & 75.27 & 87.68 & \bf 90.07\\
Soybean Clean & 50 & 564 & 74.29 & 72.87 & 64.00 & 75.18 & 81.91 & 78.90 & 86.70 & \bf 94.68\\
Wheat & 50 & 162 & 99.38 & 99.38 & 98.15 & 99.38 & \bf 100.00 & \bf 100.00 & \bf 100.00 & 99.38\\
Woods & 50 & 1244 & 92.44 & 79.98 & 90.03 & 92.04 & 87.70 & 94.21 & \bf 99.92 & 95.90\\
Buildings Grass Trees Drives & 50 & 330 & 83.94 & 67.58 & 81.52 & 89.39 & 93.33 & 92.73 & \bf 98.18 & 97.58\\
Stone Steel Towers & 50 & 45 & \bf 100.00 & \bf 100.00 & \bf 100.00 & \bf 100.00 & \bf 100.00 & 97.78 & \bf 100.00 & \bf 100.00\\
Alfalfa & 15 & 39 & 74.36 & 94.87 & 71.79 & 92.31 & \bf 100.00 & 94.87 & 97.44 & 94.87\\
Grass Pasture Mowed & 15 & 11 & \bf 100.00 & \bf 100.00 & \bf 100.00 & \bf 100.00 & 90.91 & \bf 100.00 & \bf 100.00 & \bf 100.00\\
Oats & 15 & 5 & \bf 100.00 & \bf 100.00 & \bf 100.00 & \bf 100.00 & \bf 100.00 & \bf 100.00 & \bf 100.00 & \bf 100.00\\
\hline
\multicolumn{3}{c|}{OA (\%)} & 80.07 & 75.46 & 78.02 & 82.79 & 77.56 & 86.51 & 90.64 & \bf 93.08\\
\hline
\multicolumn{3}{c|}{AA (\%)} & 86.97 & 85.31 & 84.30 & 90.03 & 88.93 & 92.33 & 94.88 & \bf 96.31\\
\hline
\multicolumn{3}{c|}{Kappa (\%)} & 77.43 & 72.13 & 75.11 & 80.44 & 74.65 & 84.63 & 89.27 & \bf 92.08\\
\hline \hline
\end{tabular}
}
\label{tab:ip result}
\end{table*}

\begin{table*}[!t]
\centering
\caption{Class-Specific and Overall Classification Accuracy (\%) of Different Methods on the Berlin Dataset}
\resizebox{1\textwidth}{!}{
\begin{tabular}{c||c|c|c|c|c|c|c|c|c|c}
\hline \hline
\multirow{2}{*}{Class Name} & \multirow{2}{*}{\makecell{Training\\Samples}} & \multirow{2}{*}{\makecell{Test\\Samples}} & \multicolumn{4}{c|}{Classic Backbone Networks} & \multicolumn{4}{c}{Hybrid Backbone Networks}\\
\cline{4-11}
~ & ~ & ~ & 2-D CNN\cite{paoletti2019deep}  & 3-D CNN\cite{chen2016deep}  & GRU\cite{pan2020spectral}  & ViT\cite{vaswani2017attention}  & MorphConv\cite{roy2021morphological}  & SSFTT\cite{sun2022spectral}  & WFCG\cite{dong2022weighted}  & DSNet\\
\hline
Forest & 443 & 54511 & 77.16 & 73.10 & 75.55 & 66.74 & 55.24 & 47.84 & \bf 76.46 & 75.44\\
Residential Area & 423 & 268219 & 66.33 & 55.27 & 57.60 & 57.86 & 72.47 & 73.58 & 68.80 & \bf 74.69\\
Industrial Area & 499 & 19067 & 21.37 & 43.92 & \bf 73.13 & 55.85 & 26.25 & 30.59 & 42.06 & 54.91\\
Low Plants & 376 & 58906 & 69.11 & 68.76 & 71.70 & 76.25 & 74.27 & \bf 84.73 & 67.58 & 81.79\\
Soil & 331 & 17095 & 68.15 & 84.05 & 82.95 & 80.04 & 84.88 & 92.20 & \bf 97.91 & 74.19\\
Allotment & 280 & 13025 & 58.52 & 32.48 & 22.30 & 64.17 & 61.21 & 49.21 & \bf 72.80 & 63.71\\
Commercial Area & 298 & 24526 & 47.47 & 30.50 & 10.64 & 29.90 & 49.90 & 39.25 & \bf 53.21 & 30.56\\
Water & 170 & 6502 & 65.70 & 58.27 & 67.24 & \bf 78.05 & 65.47 & 63.63 & 71.16 & 67.44\\

\hline
\multicolumn{3}{c|}{OA (\%)} & 64.94 & 57.77 & 59.75 & 60.97 & 67.60 & 68.23 & 68.83 & \bf 72.09\\
\hline
\multicolumn{3}{c|}{AA (\%)} & 59.23 & 55.80 & 57.64 & 63.61 & 61.21 & 60.13 & \bf 68.75 & 65.34\\
\hline
\multicolumn{3}{c|}{Kappa (\%)} & 50.06 & 42.84 & 45.02 & 47.28 & 53.38 & 53.73 & 56.48 & \bf 59.27\\
\hline \hline
\end{tabular}
}
\label{tab:berlin result}
\end{table*}

\begin{table*}[!t]
\centering
\caption{Class-Specific and Overall Classification Accuracy (\%) of Different Methods on the Augsburg Dataset}
\resizebox{1\textwidth}{!}{
\begin{tabular}{c||c|c|c|c|c|c|c|c|c|c}
\hline \hline
\multirow{2}{*}{Class Name} & \multirow{2}{*}{\makecell{Training\\Samples}} & \multirow{2}{*}{\makecell{Test\\Samples}} & \multicolumn{4}{c|}{Classic Backbone Networks} & \multicolumn{4}{c}{Hybrid Backbone Networks}\\
\cline{4-11}
~ & ~ & ~ & 2-D CNN\cite{paoletti2019deep}  & 3-D CNN\cite{chen2016deep}  & GRU\cite{pan2020spectral}  & ViT\cite{vaswani2017attention}  & MorphConv\cite{roy2021morphological}  & SSFTT\cite{sun2022spectral}  & WFCG\cite{dong2022weighted}  & DSNet \\
\hline
Forest & 146 & 13361 & 85.76 & 93.02 & 81.10 & 82.44 & 93.21 & \bf 95.52 & 91.65 & 93.19\\
Residential Area & 264 & 30065 & 87.44 & 92.68 & 84.51 & 88.77 & 95.83 & 96.39 & \bf 98.96 & 95.85\\
Industrial Area & 21 & 3830 & 69.56 & 21.46 & \bf 84.41 & 79.27 & 32.69 & 25.93 & 50.97 & 62.85\\
Low Plants & 248 & 26609 & 81.39 & 78.29 & 82.25 & 80.51 & 88.75 & 87.02 & 87.33 & \bf 91.28\\
Allotment & 52 & 523 & 41.68 & 58.70 & 32.10 & 33.35 & 63.67 & 45.70 & 52.35 & \bf 66.92\\
Commercial Area & 7 & 1638 & 14.10 & 9.46 & 5.56 & 12.39 & \bf 15.81 & 11.78 & 1.05 & 11.05\\
Water & 23 & 1507 & 35.83 & 9.36 & 19.18 & 19.04 & 17.25 & 26.28 & 33.92 & \bf 49.24\\

\hline
\multicolumn{3}{c|}{OA (\%)} & 81.33 & 80.67 & 79.85 & 81.04 & 86.39 & 86.05 & 87.66 & \bf 89.30\\
\hline
\multicolumn{3}{c|}{AA (\%)} & 59.40 & 51.85 & 55.59 & 56.58 & 58.17 & 55.52 & 59.46 & \bf 67.20\\
\hline
\multicolumn{3}{c|}{Kappa (\%)} & 73.27 & 72.01 & 71.41 & 72.96 & 80.25 & 79.59 & 82.07 & \bf 84.62\\
\hline \hline
\end{tabular}
}
\label{tab:augsburg result}
\end{table*}

\section{Experiments}
In this section, seven mainstream classification approaches are adopted to conduct comparative experiments on three HSI baseline datasets, including 2-D CNN\cite{paoletti2019deep}, 3-D CNN\cite{chen2016deep}, GRU\cite{pan2020spectral}, vision transformer (ViT)\cite{vaswani2017attention}, morphological convolutional network (MorphConv)\cite{roy2021morphological}, spectral-spatial feature tokenization transformer (SSFTT)\cite{sun2022spectral}, and weighted feature fusion of CNN and graph attention network (WFCG)\cite{dong2022weighted}. The comparison classification methods cover the classic backbone networks based on CNN, RNN, and Transformer, and a series of hybrid backbone networks to verify the effectiveness of the proposed DSNet considering subpixel information within the HSI. The overall accuracy (OA), average accuracy (AA) and Kappa coefficient (Kappa) are employed to make a quantitative performance comparison for HSI classification tasks.

\subsection{Dataset Description and Experimental Setup}
\subsubsection{Indian Pines Dataset}
The Indian Pines dataset was captured by the AVIRIS sensor \cite{green1998imaging} in June 1992, with 145 $\times$ 145 pixels and 20 m spatial resolution. It contains 220 spectral bands ranging from 400 to 2500 nm. After removing the bands covering the range of water absorption and noise, only 200 bands are remained and there are 16 classes on this dataset, see Table \ref{tab:ip result}.
\subsubsection{Berlin Dataset}
The Berlin dataset was gathered from the EnMAP data\footnote{http://doi.org/10.5880/enmap.2016.002.} and described the Berlin urban and its rural neighboring area at 30 m spatial resolution. It consists of 1723 $\times$ 476 pixels with 244 spectral bands ranging from 400 to 2500 nm. The GT data is generated by using the OpenStreetMap data and contains 8 land cover classes, as detailed in Table \ref{tab:berlin result}.
\subsubsection{Augsburg Dataset}
The Augsburg dataset was collected by the HySpex sensor\cite{baumgartner2012characterisation} over the city of Augsburg, Germany. This dataset has a spatial resolution of 30 m with the size of 332 $\times$ 485 pixels and 180 spectral bands covering the wavelength range from 400 to 2500 nm. The HSI image is divided into 7 classes, as shown in Table \ref{tab:augsburg result}.

\begin{figure*}[!t]
	  \centering
		\subfigure{
			\includegraphics[width=1\textwidth]{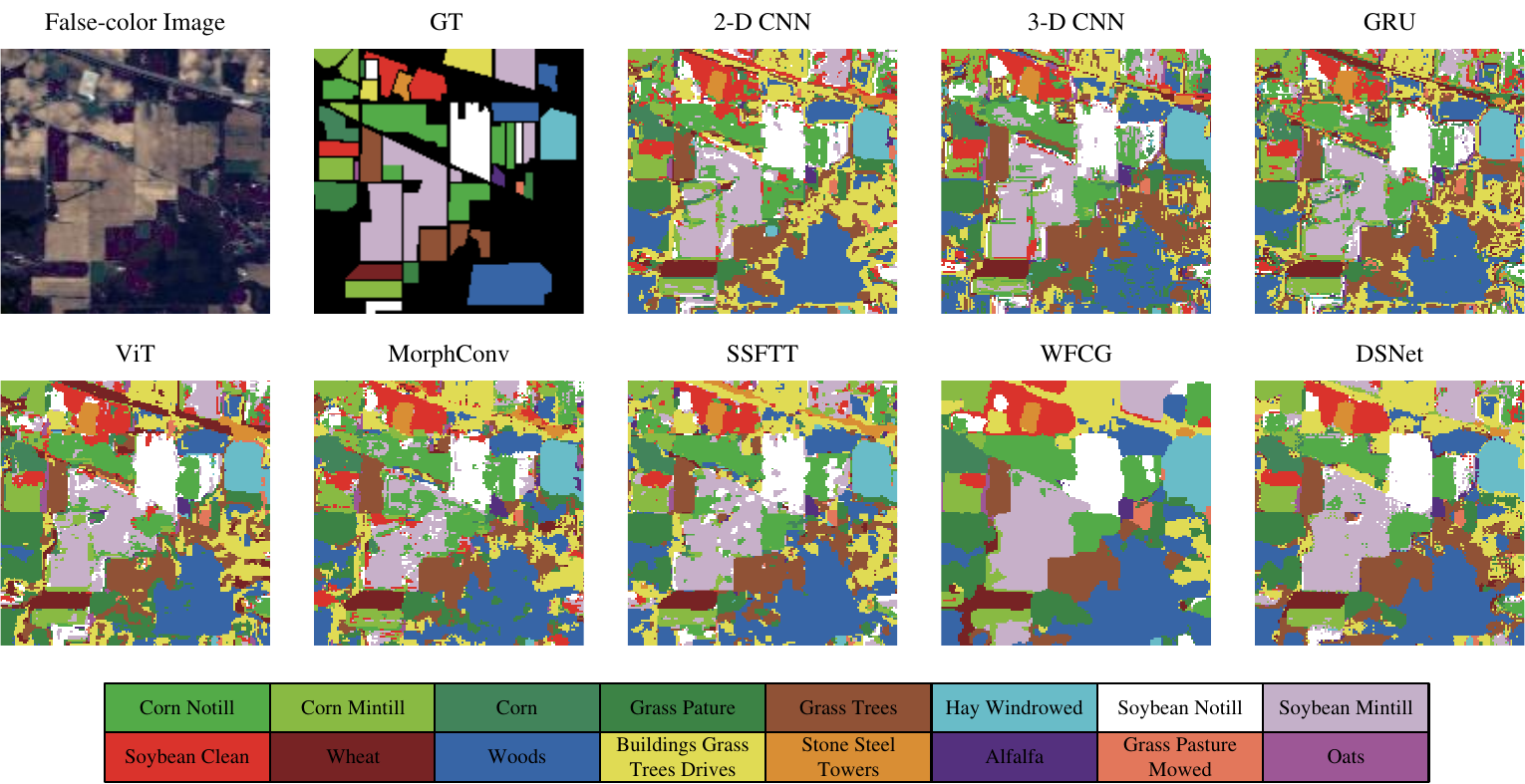}
		}
        \caption{False-color image, GT and classification maps obtained by different methods on the Indian Pines dataset.}
\label{fig: ip result}
\end{figure*}

\begin{figure*}[!t]
	  \centering
		\subfigure{
			\includegraphics[width=1\textwidth]{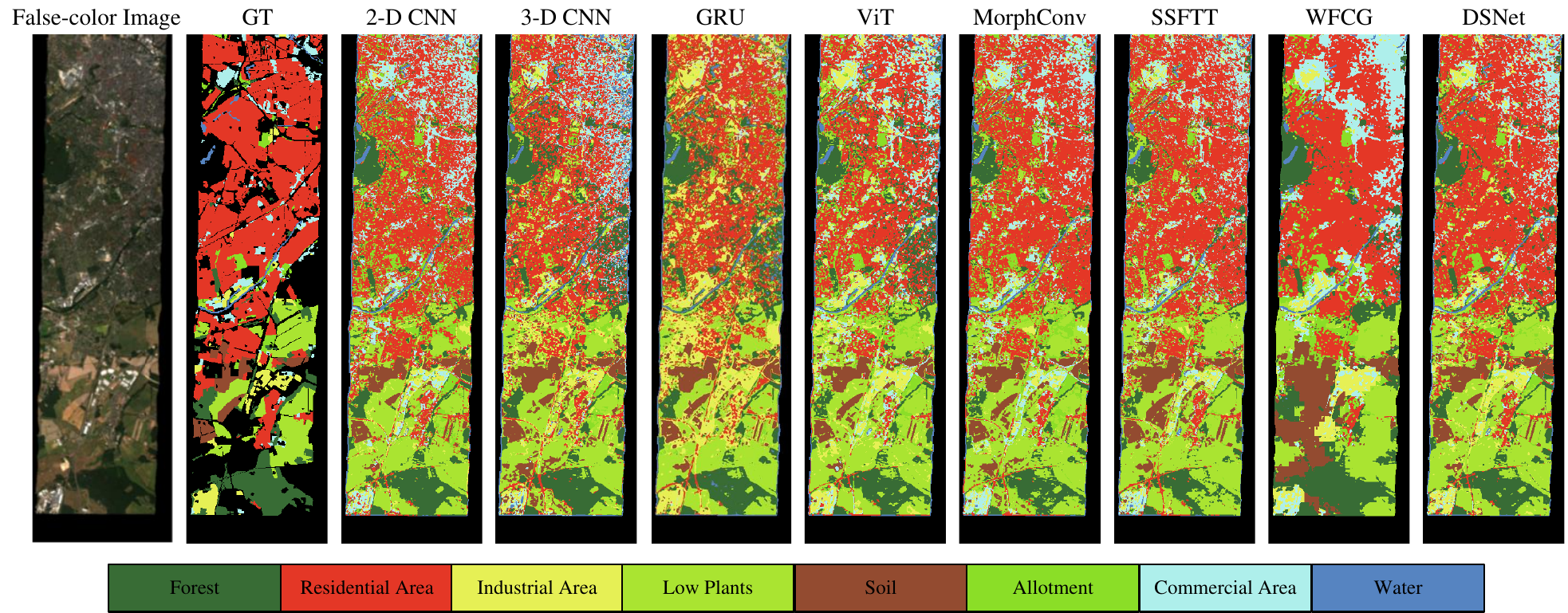}
		}
        \caption{False-color image, GT and classification maps obtained by different methods on the Berlin dataset.}
\label{fig: berlin result}
\end{figure*}

\begin{figure*}[!t]
	  \centering
		\subfigure{
			\includegraphics[width=1\textwidth]{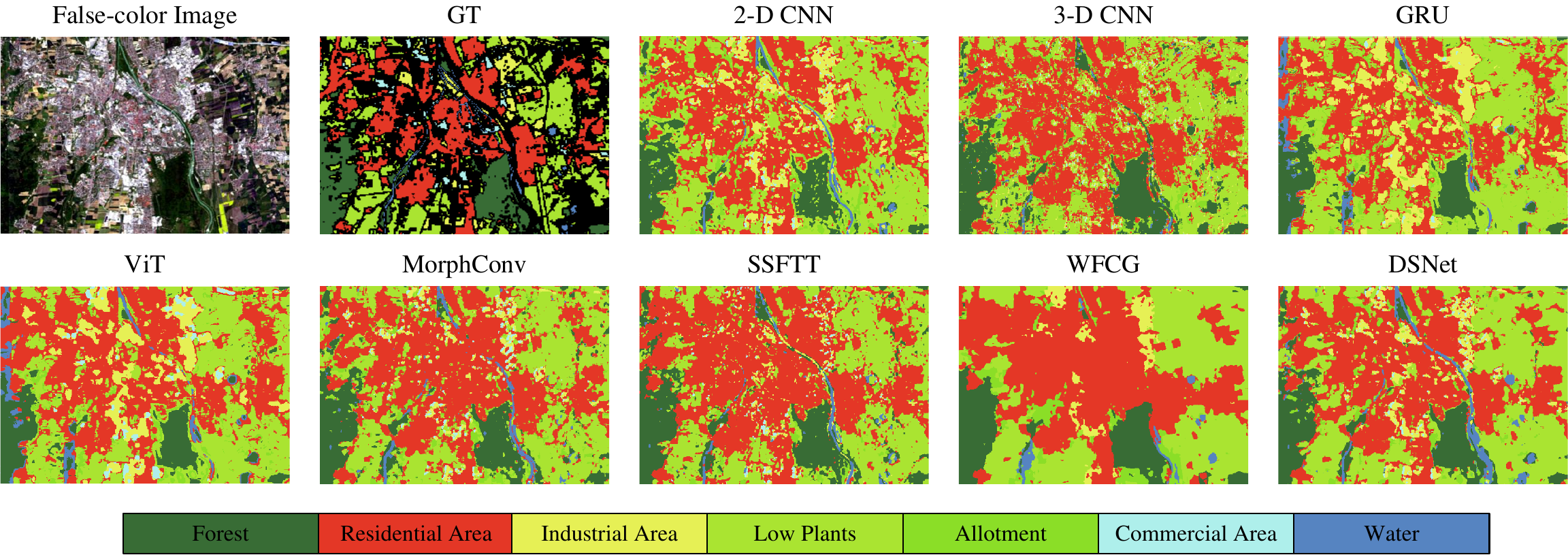}
            }
        \caption{False-color image, GT and classification maps obtained by different methods on the Augsburg dataset.}
\label{fig: augsburg result}
\end{figure*}

\begin{table*}[!t]
    \centering
    \caption{The execution time (in seconds) of one epoch training in different DL-based HSI classification methods.}
    \resizebox{0.9\textwidth}{!}{
    \begin{tabular}{c|c|c|c|c|c|c|c|c}
        \hline \hline
        Datasets & 2-D CNN\cite{paoletti2019deep}  & 3-D CNN\cite{chen2016deep}  & GRU\cite{pan2020spectral}  & ViT\cite{vaswani2017attention}  & MorphConv\cite{roy2021morphological}  & SSFTT\cite{sun2022spectral}  & WFCG\cite{dong2022weighted}  & DSNet \\
         \hline
         Indian Pines & 0.57 & 4.99 & 1.14 & 2.14 & 4.39 & 0.96 & \bf 0.30 & 0.75\\
         Berlin & \bf 28.93 & 304.04 & 54.87 & 117.63 & 205.64 & 41.40 & 33.76 & 39.78\\
         Augsburg & 3.63 & 34.55 & 7.26 & 12.51 & 31.80 & 6.42 & \bf 3.24 & 5.86\\
        \hline \hline
    \end{tabular}
    }
    \label{tab:computation}
\end{table*}

The hardware environment used for experiments is NVIDIA GTX 1080Ti 11-GB GPU. For a fair comparison, all experiments in this paper are carried out with the PyTorch framework and the comparison algorithms adopt the same hyperparameters in the original paper. Note that, different HSI classification methods adopt the consistent training and test samples on three datasets to fairly evaluate the classification performance. The Adam optimizer is adopted in DSNet to update the network parameter with a mini-batch size of 64 on all datasets. The learning rate is empirically set to $1e-3$ and decays by multiplying a factor of 0.9 after each 50 epoch. For all adopted HSI classification methods, the input is set as patch size of 7 $\times$ 7 on Indian Pines and Augsburg datasets, and 5 $\times$ 5 on Berlin dataset. The training process will be stopped after 500 epochs.

\begin{figure*}[!t]
	  \centering
		\subfigure[]{
			\includegraphics[width=0.28\textwidth]{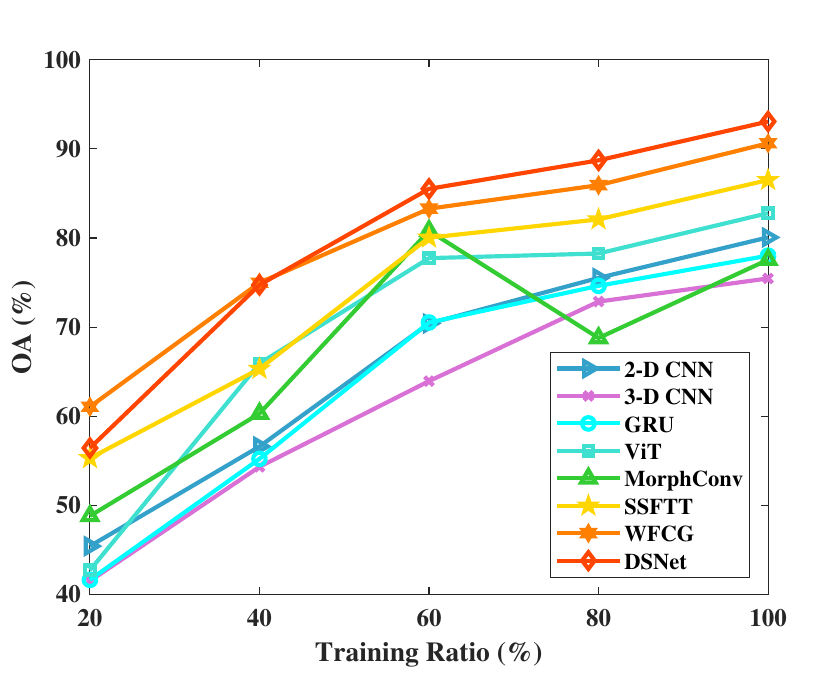}
            \label{fig:ratio_ip}
		}\qquad
		\subfigure[]{
			\includegraphics[width=0.28\textwidth]{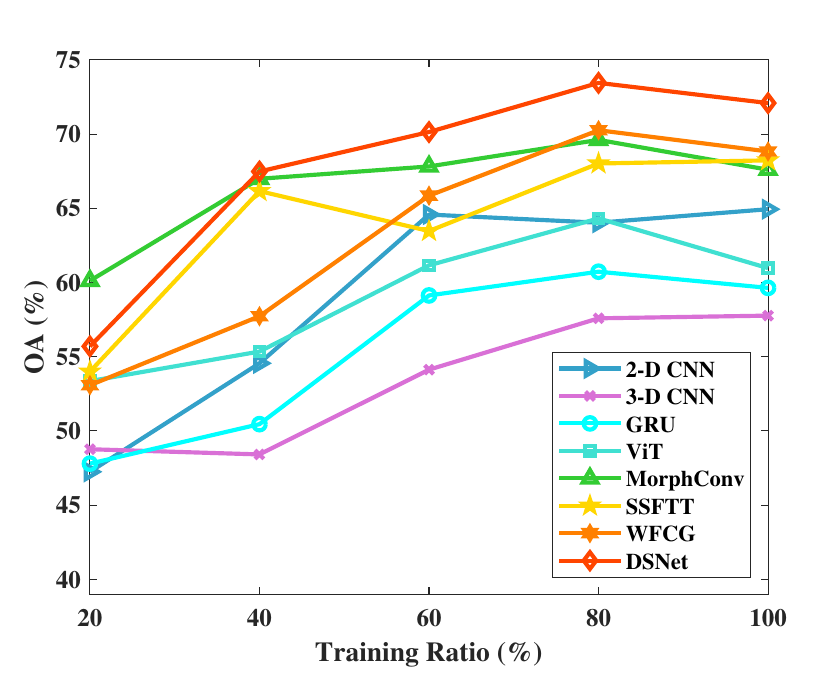}
            \label{fig:ratio_berlin}
		}\qquad
            \subfigure[]{
			\includegraphics[width=0.28\textwidth]{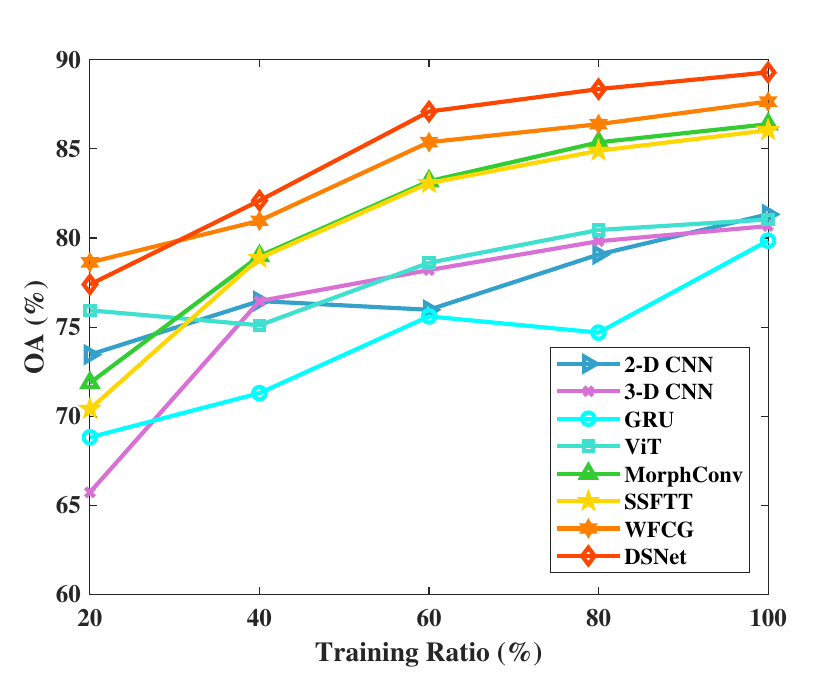}
            \label{fig:ratio_augsburg}
		}\qquad
         \caption{The classification performance of each method with different ratios of training samples on three datasets. (a) Indian Pines. (b) Berlin. (c) Augsburg.}
\label{fig:ratio}
\end{figure*}

\subsection{Comparison with Current Methods}
Tables \ref{tab:ip result} to \ref{tab:augsburg result} report the adopted training and test samples, and the quantitative results of different methods (i.e., OA (\%), AA (\%), and Kappa (\%)) on Indian Pines, Berlin and Augsburg datasets. The corresponding false-color image, GT and classification maps obtained by different methods are displayed in Figs. \ref{fig: ip result}-\ref{fig: augsburg result}. In the four classic backbone networks, 2-D CNN and ViT can obtain a better classification accuracy compared with 3-D CNN and GRU on three datasets, owing to the effective use of spatial and spectral information. The SSFTT method, due to the combination of the structural characteristics of hybrid CNN and Transformer, realizes the joint extraction process of detailed spectral-spatial features in different scenes. It successively obtains 86.51\%, 68.23\% and 86.05\% OA on three HSI classification datasets, which significantly exceeds four classic backbone networks. The WFCG method achieves mutual complementation of information by combining CNN and graph attention network to capture long-range information for HSI classification tasks, and further obtain better classification results than SSFTT. It shows approximately 4.13\%, 0.6\% and 1.61\% OA improvement over SSFTT on different dataset and the AA indicator in WFCG has been greatly improved. As can be observed from the classification results presented in Tables \ref{tab:ip result}- \ref{tab:augsburg result}, the proposed DSNet method can have the best classification performance compared with both classic backbone networks and other hybrid backbone networks. Since the deep AE unmixing module can automatically extract subpixel-level abundance maps in an unsupervised manner, DSNet can efficiently explore intrinsic high-dimensional features at the subpixel and pixel level, and further help the classifier network realize the classification process with strong robustness. In addition, it can be seen from Figs. \ref{fig: ip result}-\ref{fig: augsburg result} that DSNet achieves more accurate visualization results and preserves more spatial structures on different classes, which demonstrates the effectiveness and superiority of introducing subpixel information in the CNN-based classifier network and illustrates its outstanding classification performance.

To display the computational complexity of various HSI classification approaches, the one epoch training time on three datasets are listed in Table \ref{tab:computation}. Note that, all computational experiments are run on the same computer. It can be seen that WFCG consumes the least time on Indian Pines and Augsburg datasets, while it cannot achieve computational efficiency on large datasets, such as Berlin. The main reason is that the graph-based architecture in WFCG makes the calculation cost unbearable on large datasets, although the extracted superpixels slightly alleviate this computational burden. In addition, the architecture of self-attention and 3-D convolution enables the entire model to have high computational complexity, so that the related classification methods, i.e., 3-D CNN, MorphConv and ViT, have a large amount of execution time in the training phase. Since SSFTT treats the encoded low-dimensional features as input for self-attention calculation instead of entire HSI, it can obtain considerable computing time compared with ViT and GRU. Owing to the introduction of deep AE unmixing architecture, DSNet requires the dual-branch network training and is not superior to 2-D CNN and WFCG in terms of computational efficiency, but it almost consistently outperforms other classic and hybrid backbone networks under various HSI classification datasets. Overall, the computational cost of DSNet is acceptable for all HSI classification datasets.

\begin{table}[!t]
    \centering
    \caption{Ablation comparison of each variant in DSNet.}
    \resizebox{0.5\textwidth}{!}{
    \begin{tabular}{ccc|c|c|c|c}
        \hline \hline
        \multicolumn{3}{c|}{Module} & \multirow{2}{*}{Metrics} & \multicolumn{3}{c}{Datasets}\\
        \cline{1-3} \cline{5-7}
         Subpixel Fusion & Linear Decoder & Nonlinear Decoder & & Indian Pines & Berlin & Augsburg \\
         \hline
        \multirow{3}{*}{\xmark} & \multirow{3}{*}{\cmark} & \multirow{3}{*}{\xmark} & OA (\%) & 87.88 & 65.57 & 86.31\\
        & & & AA (\%) & 94.22 & 55.01 & 63.31\\
        & & & Kappa (\%) & 86.17 & 50.29 & 80.42\\
        \hline
        \multirow{3}{*}{\xmark} & \multirow{3}{*}{\xmark} & \multirow{3}{*}{\cmark} & OA (\%) & 89.20 & 66.81 & 87.40\\
        & & & AA (\%) & 94.87 & 63.58 & 63.87\\
        & & & Kappa (\%) & 87.54 & 53.59 & 81.90\\
        \hline
        \multirow{3}{*}{\cmark} & \multirow{3}{*}{\cmark} & \multirow{3}{*}{\xmark} & OA (\%) & 91.11 & 69.96 & 88.04\\
        & & & AA (\%) &  95.77 & 63.51 & 66.59\\
        & & & Kappa (\%) & 89.84 & 56.37 & 82.98\\
        \hline
        \multirow{3}{*}{\cmark} & \multirow{3}{*}{\xmark} & \multirow{3}{*}{\cmark} & OA (\%) & \bf 93.08 & \bf 72.09 & \bf 89.30\\
        & & & AA (\%) & \bf 96.31 & \bf 65.34 & \bf 67.20\\
        & & & Kappa (\%) & \bf 92.08 & \bf 59.27 & \bf 84.62\\
        
        \hline \hline
    \end{tabular}
    }
    \label{tab:ablation}
\end{table}

\subsection{Performance on Different Ratios of Training Samples}
In this section, we investigate the classification performance of the proposed DSNet method and the other competitors under different ratios of training samples.    Fig. \ref{fig:ratio} displays the corresponding experimental results on different HSI classification datasets. In general, the main trends of OA of all methods on three datasets increase with the ratio of training samples. When a smaller training ratio of 20\% is reached, DSNet performs less well than some hybrid backbone networks, such as WFCG and MorphConv. The main reason is that the deep AE unmixing network and the CNN-based classifier network use the same input patches, and the unmixing part requires a certain number of training samples to extract stable subpixel information in an unsupervised manner. As the ratio of training samples increases and exceeds 20\%, the obtained abundance information learned from the unmixing part becomes more complete and thus the corresponding classification performance of DSNet can significantly outperform other state-of-the-art classification methods. These results demonstrate that a certain proportion of training samples can promote the reliable generation of subpixel information, and further improve the effectiveness and stability of DSNet for HSI classification tasks.

\subsection{Model Analysis}
\subsubsection{Ablation Study}
To validate the essentiality of the proposed DSNet method, the ablation study on different network modules is investigated to determine the contribution of individual elements in this section, as listed in Table \ref{tab:ablation}. For a fair comparison, the hyperparameter settings under different network configurations are consistent and three evaluation indicators are adopted to conduct comparative analysis of classification performance on three datasets. The two primary parts of DSNet are considered in the ablation study: the subpixel fusion module and different AE mixture mechanism including linear decoder and nonlinear decoder. It is evident that considering only nonlinear decoder in the deep AE unmixing network can increase the OA by 1.32\%, 1.24\% and 1.09\% than considering only linear decoder on the Indian Pines, Berlin and Augsburg datasets, respectively. When the subpixel fusion module is integrated, this OA improvement is further increased by 1.97\%, 2.13\%, 1.26\% on three HSI classification datasets. The main reason is that adopting nonlinear decoder guarantees the general mixing modeling mechanism with physically nonlinear properties, so that the reconstruction process in the deep AE unmixing network greatly avoids the loss of detailed information and inherent physical characteristics related to different ground objects. Due to the difference between abundance maps and class-wise representation in spatial dimension, the subpixel fusion module provides a simple and effective strategy to fuse subpixel and pixel features, further ensuring the joint training of the deep AE unmixing network and the CNN-based classifier network. It can be seen from Table \ref{tab:ablation}
that the introduction of subpixel fusion module can exceed the baseline respectively by 3.23\%, 4.38\% and 1.73\% under the linear decoder setting, and by 3.88\%, 5.28\% and 1.90\% under the nonlinear decoder setting, which further demonstrates the effectiveness and superiority of the subpixel fusion module in the proposed DSNet. In addition, the collaboration of subpixel fusion module and nonlinear decoder module can significantly improve the classification performance in three evaluation indicators, making DSNet easier to explore and distinguish different classes in actual scenarios. 

\subsubsection{Parameter Tuning}
For the proposed method, there are two main tuning parameters, i.e., the number of decoder layers and $\lambda$ value, which control the nonlinear mixture degree of the deep AE unmixing network and fusion scaling ratio of the two branches, respectively. For the number of decoder layers, we set different numbers of decoder layers to evaluate the classification performance of the proposed DSNet as shown in Table \ref{tab:nonlinear layer}. As can be seen in Table \ref{tab:nonlinear layer}, the classification results are relatively stable with regard to the variation of the number of decoder layers. As the number of decoder layers increases, DSNet can always acquire stable subpixel information with physically nonlinear properties and ensure higher classification accuracy than several state-of-the-art classification methods. However, when the number of decoder layers is set to 5, the excessive number of layers in the unmixing network affects the training process of the CNN-based classifier network branch, making DSNet difficult to separate categories of different ground objects. Thus, this inspires us to set the number of decoder layers from 1 to 4. To reduce the hyperparameter optimization pressure, the number of decoder layers on all datasets is uniformly set to 2.

\begin{figure}[!t]
	  \centering
		\subfigure{
			\includegraphics[width=0.43\textwidth]{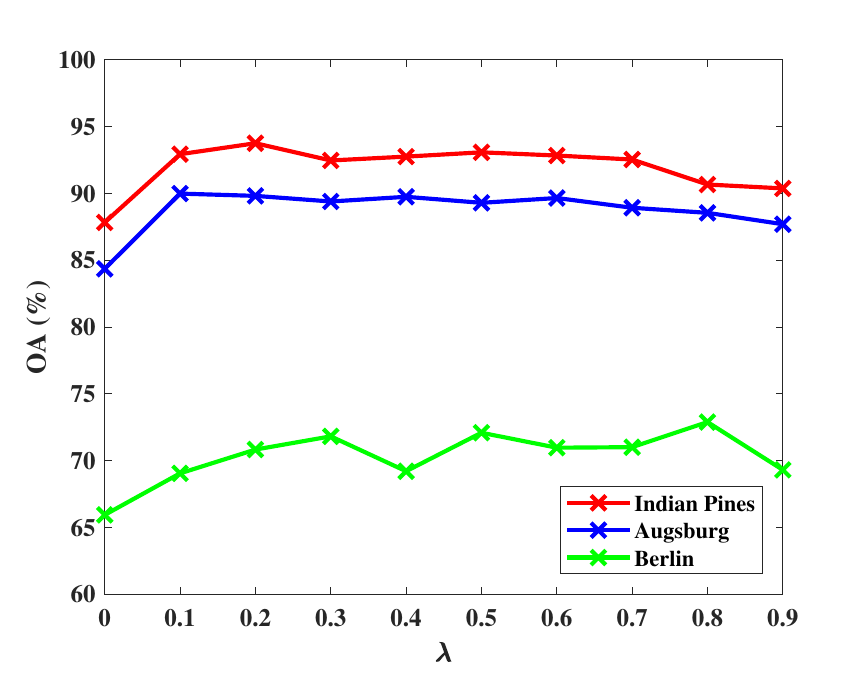}
		}
        \caption{Classification performance analysis between OA and various $\lambda$ values on different datasets.}
\label{fig: lambda}
\end{figure}

For the fusion coefficient, we set $\lambda$ in the range of [0, 0.9] with an interval of 0.1 to investigate the effect of parameters of $\lambda$. Fig. \ref{fig: lambda} illustrates the obtained results in terms of OA with different $\lambda$ values on three HSI classification datasets. When $\lambda$ is set to 0, it means that unmixing part has no effect and the proposed DSNet obtains the worst classification performance on all datasets. As the value of $\lambda$ increases, the deep AE unmixing network begins to provide a certain proportion of subpixel-level abundance information for the CNN-based classifier network, so that the classification accuracy of DSNet changes steadily with the value of $\lambda$ on three datasets. It can be seen from Fig. \ref{fig: lambda} that the best OA is achieved when $\lambda$ is equal to 0.2 on the Indian Pines dataset, 0.8 on the Berlin dataset and 0.1 on the Augsburg dataset. When $\lambda$ is set to 0.9, DSNet overemphasizes the function of the unmixing network and further exhibits the degradation of classification performance. Overall, different $\lambda$ settings show the same trend in various datasets, and it can be proven that introducing certain subpixel information can help improve the classification accuracy of DSNet. In all experiments, the value of $\lambda$ is uniformly fixed to 0.5 for a fair comparison.

\begin{table}[!t]
    \centering
    \caption{Comparison of HSI Classification Performance for Different Numbers of Decoder Layers in the Deep AE Unmixing Network.}
    \resizebox{0.45\textwidth}{!}{
    \begin{tabular}{c|c|c|c|c|c|c}
        \hline \hline
        \multirow{2}{*}{Datasets} & \multirow{2}{*}{Metrics} & \multicolumn{4}{c}{Number of Decoder Layers $K$}\\
        \cline{3-7}
         & & 1 & 2 & 3 & 4 & 5 \\
         \hline
         \multirow{3}{*}{Indian Pines} & OA (\%) & 92.14 & \bf 93.08 & 92.31 & 91.86 & 91.86\\
         & AA (\%) & 96.38 & 96.31 & 96.45 & \bf 96.82 & 96.58\\
         & Kappa (\%) & 91.27 & \bf 92.08 & 91.15 & 90.65 & 90.63\\
         \hline
         \multirow{3}{*}{Berlin} & OA (\%) & 70.85 & \bf 72.09 & 71.22 & 71.59 & 71.34\\
         & AA (\%) & 64.31 & 65.34 & \bf 65.41 & 63.88 & 60.85\\
         & Kappa (\%) & 56.86 & \bf 59.27 & 58.24 & 58.60 & 57.29\\
         \hline
         \multirow{3}{*}{Augsburg} & OA (\%) & 88.98 & 89.30 & 89.01 & \bf 91.04 & 90.08\\
         & AA (\%) & 66.52 & \bf 67.20 & 64.88 & 65.74 & 61.11\\
         & Kappa (\%) & 84.27 & 84.62 & 84.15 & \bf 87.17 & 85.64\\
        \hline \hline
    \end{tabular}
    }
    \label{tab:nonlinear layer}
\end{table}

\subsubsection{Visual Evaluation}
\begin{figure}[!t]
	  \centering
		\subfigure[]{
			\includegraphics[width=0.21\textwidth]{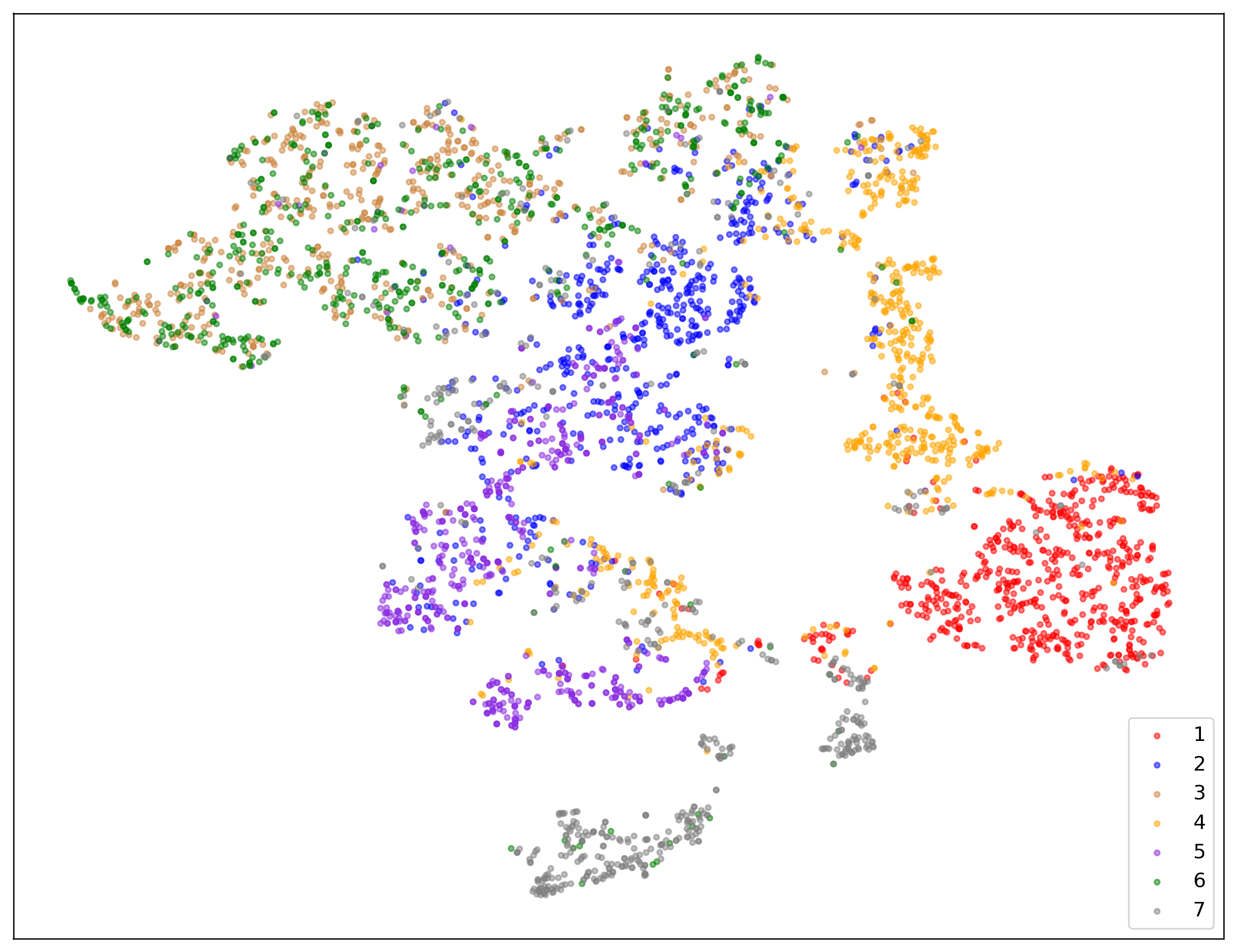}
            \label{fig:conv2d}
		}\qquad
		\subfigure[]{
			\includegraphics[width=0.21\textwidth]{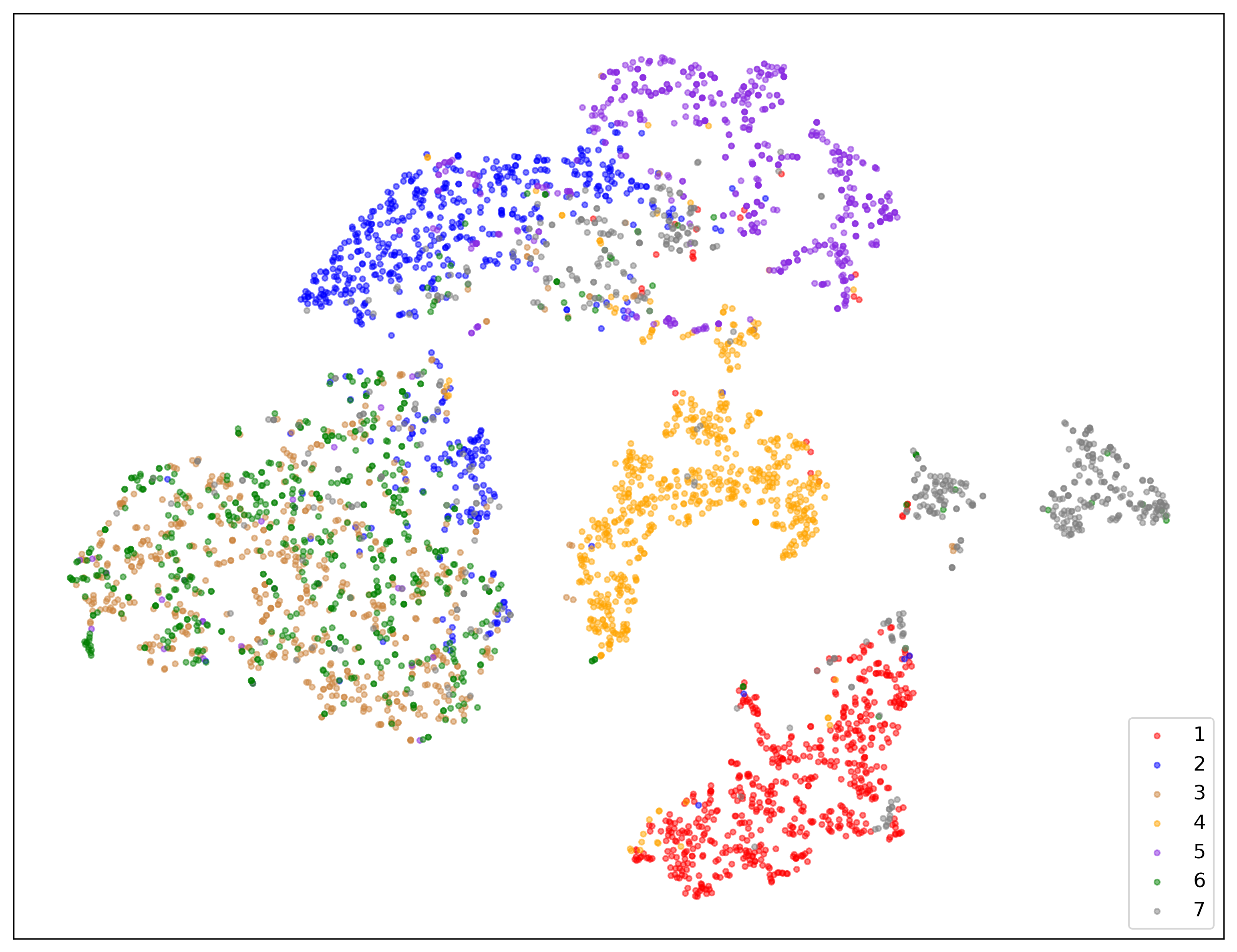}
            \label{fig:conv2d_unmix}
		}\qquad
         \caption{Visualization of class features obtained by only CNN-based classifier network and DSNet. (a) Class features output by CNN-based classifier network. (b) Class features output by DSNet that considers both CNN-based classifier network and deep AE unmixing network.}
\label{fig:tsne}
\end{figure}

As a data visualization technique, the t-distributed stochastic neighbor embedding (t-SNE) \cite{van2008visualizing} is adopted to provide better visualizations for investigating the class separability performance of different classification models. Fig. \ref{fig:tsne} displays the distribution visualization of class features obtained by only CNN-based classifier network and DSNet on the Augsburg dataset. It is apparent that the feature distribution of only CNN-based classifier network is extremely scattered and several classes have significant feature distribution overlap. For example, except for the first class, other classes do not have the tendency to form crowded points together owing to the limited feature learning ability of the CNN-based classifier network at the pixel level. By designing the deep AE unmixing network, DSNet contains richer information of different classes from both subpixel-level and pixel-level features. As a result, many classes become more separated and the distributions of the same category become more compact, such as the 4-th class, the 5-th class and 7-th class, which further demonstrates that the proposed DSNet can achieve more reliable decision boundaries for class label distribution.

\begin{figure}[!t]
	  \centering
		\subfigure{
			\includegraphics[width=0.5\textwidth]{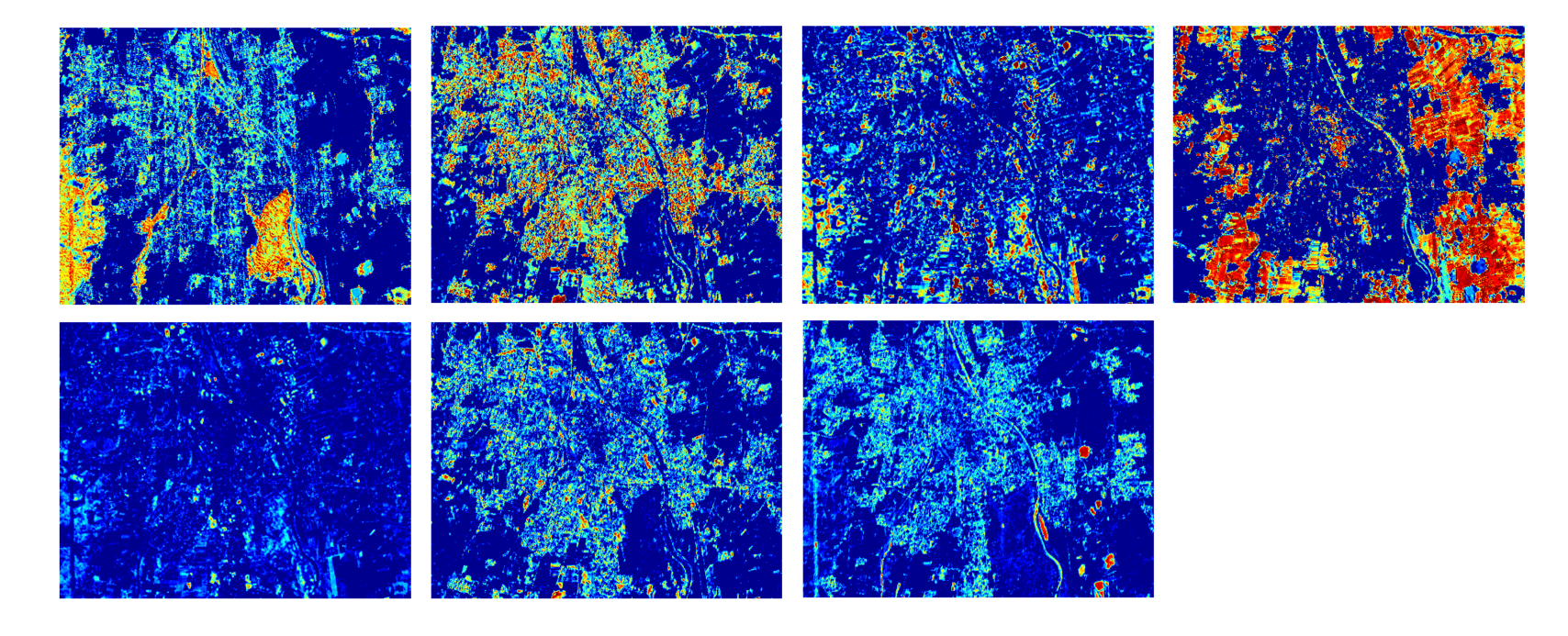}
		}
        \caption{Visualization of the extracted abundance maps obtained by DSNet.}
\label{fig: abundance}
\end{figure}

For illustrative purposes, the extracted abundance maps by DSNet on the Augsburg dataset are depicted in Fig. \ref{fig: abundance}. It is worth noting that the GT generation usually involves manual calibration of endmember spectra and inversion to produce abundance maps, making it unsuitable for different application scenarios in the classification task. Therefore, we tend to explore the usefulness of the extracted subpixel information based on classification performance, rather than quantitative evaluation of unmixing results. It can be clearly observed that DSNet can obtain more distinct abundance estimation results with the help of deep AE unmixing architecture and the extracted subpixel-level abundances provide rich spatial information and class-wise differences, thereby yielding a significant classification performance improvement. In addition, the designed subpixel fusion module helps the entire model achieve high-quality information fusion across pixel and subpixel features, so that the generated abundance maps are optimized from the perspective of both deep AE unmixing and CNN-based classifier network during the training process. The abundance results indicate the effectiveness of the proposed DSNet to achieve a DL-based collaborative subpixel-pixel framework for HSI classification tasks in real-world scenarios.

\section{Conclusion}
This paper proposes a dual-branch subpixel-guided  framework (DSNet) for HSI classification in the remote sensing community. We explore the relationship between unmixing and HSI classification and design a joint classification network to deal with mixed pixels to enhance HSI classification performance in actual scenarios. Unlike existing DL-based classification approaches rely on complex network architecture design, the proposed DSNet is capable of fully exploiting inherent subpixel information and designing suitable fusion strategy to improve classification accuracy from a data-driven perspective. Based on a general mixture model, the deep AE unmixing network can achieve automatic subpixel-level abundance extraction with certain physically significance in an unsupervised manner. Then, the subpixel fusion module is developed to ensure high-quality information fusion across pixel and subpixel features, which further achieves stable joint classification and facilitates better class separation. Experimental results performed on several real HSI classification datasets demonstrate the superior performance of our proposed method compared with state-of-the-art DL-based HSI classification methods. Future work will integrate more complete spectral-spatial classifier, such as the hybrid architecture of CNN and Transformer, into the AE network to further enhance its performance.

\bibliographystyle{IEEEtran}
\bibliography{HZ_ref}

\end{CJK}
\end{document}